%% file: main.tex
\documentclass{egpubl}
\usepackage{pgv2023}
 
\ConferencePaper        
\WsPaper           

\usepackage[T1]{fontenc}
\usepackage{dfadobe}  
\usepackage{amsfonts}

\newcommand{\scenery}{\textit{scenery} \cite{guenther2019scenery}}
\newcommand{\by}{$\times$}
\newcommand{\insitu}{\textit{in situ}}
\newcommand{\vdirendering}{\cite{pacific}}
\newcommand{\githuburl}{\url{https://github.com/scenerygraphics/scenery-insitu/}}
\newcommand{\sensitivity}{$\gamma$}
\newcommand{\criteria}{$\tau$}
\newcommand{\processing}{\textbf{PE}}
\newcommand{\processings}{{\processing}s}
\newcommand{\superseg}[2]{\ensuremath{\mathbb{S}^{#1}_{#2}}}
\newcommand{\subsuperseg}[2]{sub-\ensuremath{\superseg{#1}{#2}}}
\newcommand{\supseglist}[1]{\ensuremath{\mathbb{L}_{#1}}}
\newcommand{\front}[2]{\ensuremath{f(\mathbb{S}^{#1}_{#2})}}
\newcommand{\back}[2]{\ensuremath{b(\mathbb{S}^{#1}_{#2})}}
\newcommand{\vieworig}[0]{\ensuremath{\mathrm{V}_\mathrm{O}}}
\newcommand{\viewnew}[0]{\ensuremath{\mathrm{V}_\mathrm{N}}}

\newcommand{\numSupsegs}[0]{\ensuremath{\mathbb{N_{S}}}}
\newcommand{\numLists}[0]{\ensuremath{\mathbb{N_{L}}}}
\newcommand{\fhdres}{1920$\times$1080}

\usepackage{caption}
\usepackage{subcaption}
\usepackage{cite}  
\BibtexOrBiblatex
\electronicVersion
\PrintedOrElectronic
\ifpdf \usepackage[pdftex]{graphicx} \pdfcompresslevel=9
\else \usepackage[dvips]{graphicx} \fi
\usepackage{todonotes}
\usepackage{egweblnk} 

\usepackage{algpseudocode}
\usepackage{algcompatible}
\usepackage[linesnumbered,ruled,vlined]{algorithm2e}
\usepackage{cuted}

\title[Parallel Compositing of VDIs]{Parallel Compositing of Volumetric Depth Images for Interactive Visualization of Distributed Volumes at High Frame Rates}

\author[Gupta et al.]{
{\parbox{\textwidth}{\centering Aryaman Gupta$^{1,2,3}$\orcid{0009-0002-3287-0601},
         Pietro Incardona$^{1,2,3}$,
         Anton Brock$^{1}$,
         Guido Reina$^{4}$\orcid{0000-0003-4127-1897},
         Steffen Frey$^{5}$\orcid{0000-0002-1872-6905},
         Stefan Gumhold$^{1}$\orcid{0000-0003-2467-5734},\\
         Ulrik G\"{u}nther$^{6,2,3}$\orcid{0000-0002-1179-8228},
        and Ivo F. Sbalzarini$^{1,2,3}$\orcid{0000-0003-4414-4340}
        }}
        \\
{\parbox{\textwidth}{\centering $^1$Technische Universit\"{a}t Dresden, Faculty of Computer Science, Dresden, Germany\\
        $^2$Max Planck Institute of Molecular Cell Biology and Genetics, Dresden, Germany\\
        $^3$Center for Systems Biology Dresden (CSBD), Dresden, Germany\\
        $^4$VISUS, University of Stuttgart, Stuttgart, Germany\\
        $^5$University of Groningen, Groningen, The Netherlands\\
        $^6$Center for Advanced Systems Understanding (CASUS), G\"{o}rlitz, Germany\\
       }
}
}

\begin{document}


\maketitle
\begin{abstract}
We present a parallel compositing algorithm for Volumetric Depth Images (VDIs) of large three-dimensional volume data.
Large distributed volume data are routinely produced in both numerical simulations and experiments, yet it remains challenging to visualize them at smooth, interactive frame rates.
VDIs are view-dependent piecewise constant representations of volume data that offer a potential solution.
They are more compact and less expensive to render than the original data.
So far, however, there is no method for generating VDIs from distributed data.
We propose an algorithm that enables this by sort-last parallel generation and compositing of VDIs with automatically chosen content-adaptive parameters.
The resulting composited VDI can then be streamed for remote display, providing responsive visualization of large, distributed volume data. 


\end{abstract}  
\section{Introduction}
Interactive volume rendering is commonly used when analyzing large scalar field data generated by scientific simulations or experimental measurement devices. Rendering at high, consistent frame rates and low latency is crucial for enabling smooth viewpoint changes and zooming, which are important for gaining depth perception and spatial understanding. Distributed compute clusters are therefore commonly used to accelerate the rendering of large data, distributing the data and parallelizing the calculations among processors. But consistently high frame rates are difficult to achieve in such a setting, due to the time-consuming raycasting procedure and the remote rendering setup, which introduces network latency.

Here, we propose the use of view-dependent piecewise-constant representations of volume data, also known as Volumetric Depth Images (VDIs) \cite{Frey}, to decouple interactive viewpoint changes and zooming from network latency and distributed volume raycasting. 
VDIs are generated by dividing the volume-rendering integral along each ray into chunks that store accumulated
color and opacity. The resulting representation can be much smaller than the volume data \cite{Frey}, can be compressed and streamed efficiently \cite{spacetime}, and recent work has shown that it can be rendered at high frame rates, providing high-fidelity approximations near the viewpoint from which it was generated \vdirendering. However, there currently exists no method for generating VDIs on distributed volume data.

We therefore present an algorithm for sort-last parallel generation of VDIs on distributed data. In this approach, VDIs are generated on each processing element (\processing) on its volume domain in parallel---we call these ``sub-VDIs''---and composited in parallel to a single VDI with load-balancing in image space. We design the present algorithm such that it can adapt to arbitrary, potentially non-convex data decompositions, as may arise, for example, in \insitu~visualization of distributed computer simulations.

We benchmark the proposed method on real-world datasets. We test the parallel compositing algorithm for accuracy and scalability, showing that it can be used to enable responsive visualization at high frame rates for large distributed volume data. We provide our implementation as part of the open-source visualization library \scenery.
In summary, we contribute the following:
\begin{itemize}
    \item We propose the use of view-dependent piecewise-constant volume representations, such as VDIs, for interactive visualization of distributed volumes at high, consistent frame rates.
    \item We provide an efficient parallel algorithm for scalable sort-last generation of VDIs over distributed data.
\end{itemize}

\section{Related Work}

Before presenting the parallel VDI generation and compositing algorithm for distributed-memory volume data, we review the state of the art and related works in relevant areas and recall the main VDI concepts.

\subsection{Distributed Volume Rendering}
Volume rendering is widely used for the visualization of 3D scalar fields. Soon after the volume raycasting algorithm was first presented by Levoy \cite{levoy1988display}, parallel volume rendering began to receive research interest \cite{neumann1933parallel, ma1993data} with the purpose of achieving interactive visualization by distributing the data and parallelizing the rendering calculations. Later progress in parallel volume rendering enabled efficient rendering at high degrees of parallelism and for large data sizes \cite{peterka2009end,howison2011hybrid}.

A commonly used strategy for parallel volume rendering is sort-last rendering \cite{sortlast,peterka2009end,howison2011hybrid}. There, the volume data are distributed among the $n$ \processings~taking part in the rendering. Each \processing~performs front-to-back raycasting on its data, producing a full-resolution sub-image. The sub-images from the different \processings~are then composited to a single image of the overall data.

Cavin et al.~\cite{cavin2005cots} provided a theoretical comparison of algorithms used for compositing the sub-images. Perhaps the most straightforward is the direct-send algorithm \cite{neumann1933parallel}, where the image is divided among the $n$ \processings~such that each  is responsible for compositing $1/n$-th of the total pixels in the final image. For this, each \processing~receives fragments of images from all other \processings , corresponding to the part of the final image that it ``owns''. Peterka et al.~\cite{peterka2009end} used the direct-send approach to demonstrate scalability of parallel volume rendering on an IBM BlueGene/P system.

Other frequently used compositing algorithms include the binary-swap algorithm \cite{291532}, which uses a tree structure with pairs of processes communicating at every node of the tree, and the hybrid radix-k compositing algorithm \cite{peterka2009configurable}, which blends the direct-send and binary-swap approaches offering configurable parameters for optimization on different hardware architectures. Previous work has aimed to optimize compositing by dynamically scheduling communication of sub-images for better overlap with computation \cite{grosset-dynamic}, reducing communication cost using hybrid OpenMP/MPI parallelism and GPU-Direct RDMA \cite{grosset-rdma}, and by compressing image data on the GPU before compositing \cite{icetnew}. Interactive frame rates, however, still require a finely granular domain decomposition to reduce rendering times, and responsive visualization remains a challenge due to network latency between the user and the parallel rendering cluster.

\subsection{Explorable Image Representations}

For responsive remote visualization, different explorable image representations have been proposed in order to decouple user interactions from rendering. Shade et al.~\cite{layered} introduced the view-dependent Layered Depth Image (LDI) representation, storing multiple pixels along each line of sight. Stone et al.~\cite{omni} used omnidirectional stereoscopic images, rendered on remote compute clusters and reprojected locally, to enable Virtual Reality (VR) visualization of molecular dynamics simulations. These approaches, however, are limited to surface and geometry data. For reprojecting volume data, Zellmann et al.~\cite{zellmann2} used a single depth layer transmitted by the rendering server together with the color buffer, proposing various heuristics to create the depth buffer. While using a single depth value per pixel ensures small message sizes, it is not conducive to producing high-quality reprojections, as holes occur where rays do not intersect the depth layer. This has been addressed by view-dependent piecewise-constant volume representations~\cite{brady, Frey, novelview}, which produce a continuous representation of the volume by storing multiple layers with composited color and opacity in-between. One such representation is the VDI, as described in more detail in Sec.~\ref{sec:vdi_basics}.


Tikhonova et al.~\cite{tikhonova_multiple, tikhonova_coherent} and, more recently, Rapp et al.~\cite{Rapp}, proposed compact view-dependent representations that enable interactive transfer function changes. The VDI differs from these representations in that it is generated for a given transfer function and stores transfer-function classified color and opacity within each segment. This prioritizes fast rendering from new viewpoints. Rapp et al.~\cite{Rapp} also support viewpoint changes, but require slower bilinear interpolation of Lagrange multipliers.

Recent years have seen increased research interest in novel-view synthesis---using one or more images of a scene to generate images from new viewpoints---using deep-learning techniques. Mildenhall et al.~\cite{nerf} proposed the NeRF (Neural Radiation Fields) representation, training a neural network to encode a continuous volume in its weights, enabling rendering by raycasting over samples collected from the network. 
This has been extended to implicit neural representations achieving high compression ratios on large volume data~\cite{nnvolume, nnvolume2}.
While techniques have been proposed to accelerate neural rendering \cite{ngp, nnvolumefast}, dense regions in large volumes still require many samples, limiting frame rates.

Exploratory visualization of numerical simulation results was done post-hoc using the Cinema database \cite{cinema}, which stores images generated \insitu~for a range of visualization parameters, including different viewpoints. All parameter ranges, including viewpoints, however, must be specified in advance, and the database becomes large if many viewpoints are required. Our approach instead generates new VDIs at regular time intervals, which can be streamed to enable approximate rendering with full six-degrees-of-freedom viewpoint changes.

\subsection{Generation of View-Dependent Piecewise-Constant Volume Representations} \label{sec:view_dependent}

We review techniques that have been proposed for generating view-dependent piecewise-constant volume representations, including VDIs \cite{Frey}.

All view-dependent representations of volume data are generated by raycasting through the volume and decomposing the volume-rendering integral into segments, each of which containing transfer-function classified composited color and opacity. The distinguishing feature of these representations, in comparison to other techniques that compress volume data, is that they produce an exact image when rendered from the original viewpoint of generation, owing to the associativity of the {\texttt{over}} operator \cite{over} used in alpha-compositing. Rendering from deviating viewpoints involves accumulation over the segments, which is much cheaper than evaluating the full integral \cite{overview}. Close approximations to volume rendering are achieved around the viewpoint of generation \cite{Frey}.

Previous works on generating view-dependent volume representations differ in the strategies used to determine the locations and extents of the segments generated along the rays. Brady et al. \cite{brady} used constant-size segments along each ray, creating segments that contain composited color and opacity over potentially heterogeneous samples, hampering the quality of rendering from a new viewpoint. Lochmann et al. \cite{novelview} created segments of constant opacity by partitioning the total transmittance equally among the segments. This, however, does not account for potentially varying color values within the segments. Frey et al. \cite{Frey} proposed the VDI, which uses homogeneity as a criterion for segment termination, accumulating samples into a segment unless it differs from the current segment by more than a user-defined sensitivity parameter. Recent work \vdirendering~has proposed an iterative search strategy to automatically determine the sensitivity parameter. However, while VDIs can provide high-fidelity approximate renderings, there exists so far no method for generating them in parallel on distributed volume data.

\subsection{The Volumetric Depth Image (VDI)}\label{sec:vdi_basics}
Frey et al.~\cite{Frey} proposed the VDI as a view-dependent representation of volume data. They call the segments generated along each ray \textit{supersegments}. Each supersegment \superseg{}{}~is represented by its front and back faces, \front{}{} and \back{}{}, and its color and opacity, $C($\superseg{}{}$)$ and $\alpha($\superseg{}{}$)$, respectively.

\begin{figure}
\centering 
\includegraphics[width=\columnwidth]{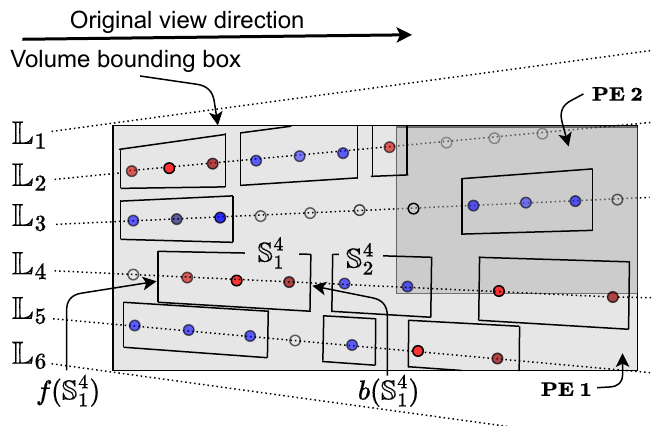}
\caption{A Volumetric Depth Image (VDI) \cite{Frey} is generated by casting rays through the volume and grouping samples (depicted by colored circles along the ray) of similar color and opacity, generating a list \supseglist{i} of up to \numSupsegs=3 (in the example of the figure) supersegments \superseg{i}{j} per ray. Each \superseg{i}{j} stores its front and back face, \front{i}{j} and \back{i}{j}, along with color and opacity accumulated in-between (Fig.~\ref{fig:dense}a). Hollow circles represent samples in empty regions. Volume data may be divided among multiple Processing Elements (\processing)~in a computer cluster (background gray levels).}
\label{fig:generation}
\end{figure}

Each ray $(x,y)$ cast into the volume creates a supersegment \textit{list} \supseglist{xy} of up to \numSupsegs~(user parameter) supersegments \superseg{xy}{j}, where $j$ is the index of the supersegment within the list (Fig.~\ref{fig:generation}). The total number of lists created, \numLists, corresponds to the viewport resolution the VDI is generated for, i.e.  \numLists$ = wh$, where $w$ is the width of the viewport and $h$ the height in pixels. 

The decomposition of the volume rendering integral into supersegments is governed by a termination criterion \criteria, which depends on a sensitivity parameter \sensitivity. Samples along each ray are merged into a supersegment until
\begin{equation}\label{eq:termination}
   \tau\, : \, \gamma < \| C(\mathbb{S})\alpha(\mathbb{S}) - C'\alpha' \|_2 \,,
\end{equation}
where $C'$ and $\alpha'$ are the color and the length-adjusted opacity of the next sample. In words, the next sample is composited into the current \superseg{}{} unless it differs from the premultiplied color of \superseg{}{} by more than \sensitivity, in which case a new \superseg{}{} is started. 
This criterion generates homogeneous \superseg{}{}, which is important for generating high-quality approximated renderings from new viewpoints.

The value of \sensitivity~that generates the most accurate VDI depends on the dataset and on the transfer function. A greedily optimal per-ray value is found automatically by iterative bisection search \vdirendering. At each iteration, a pass is performed through the volume, and the total number of times \criteria~is true is used to modify the value of \sensitivity~for the next iteration given a \numSupsegs~budget. If \criteria~was true more often than \numSupsegs , the value of \sensitivity~is increased for the next iteration, and vice versa. 
This process iterates until \sensitivity~has converged to a value that maximizes the number of supersegments generated while staying below a total of \numSupsegs . This iterative per-ray \sensitivity~optimization enables VDIs to be generated without manual intervention, accurately adapting to the data and the transfer function.

In the following, we extend the VDI concept to distributed parallel generation and compositing. More specifically, we propose a compact memory layout for VDIs in order to reduce the communication overhead during parallel compositing, we show how \sensitivity~search can be approximated without global communication, and we handle arbitrary non-convex data-domain decompositions. All design decisions are rationalized by performance measurements.

\section{Measurement Setup}
We compare and evaluate design decisions and algorithmic approaches in a uniform benchmark setup. All measurements are taken on the {\em taurus} high-performance computer of TU Dresden. Each node contains 8 NVIDIA A100-SXM4 GPUs with 40 GB of DRAM each, 2 AMD EPYC 7352 CPUs with 24 cores each, 1 TB RAM, and runs RedHat Enterprise Linux version 7.9. C++ code was compiled using GCC 10.3.0, Java code was run using OpenJDK 11.0.2, and OpenMPI version 4.1.1 was used. Processes are always distributed in a block manner
, i.e., all 8 GPUs on a node are occupied before starting to use another.

The datasets used for the measurements are described in Table \ref{table:datasets}. All of them are commonly used for the evaluation of visualization tools and algorithms and are referred to by their name or abbreviation. Volume data is decomposed across \processings~in blocks of as close to equal size and extents along the spatial dimensions as permitted by the data dimensions. Rendering of VDIs, performed to verify the quality of VDIs generated, ran on a Nvidia RTX 4090 on a remote workstation under Ubuntu 20.04.

\begin{table}[]
\centering
\begin{tabular}{|p{3cm}|p{2.8cm}|l|}
\hline
Dataset (Abbreviation)          & Dimensions                                                              & Size    \\ \hline
Kingsnake           & 1024\by1024\by795 ~8bit uint   & 795 MiB \\ \hline
Rayleigh-Taylor Instability\cite{miranda}     & 1024\by1024\by1024 16bit uint & 2 GiB   \\ \hline
Richtmyer-Meshkov\cite{richtmyer} & 2048\by2048\by1920  8bit uint  & 7.5 GiB \\ \hline
Rotating Stratified Turbulence\cite{rotstrat} (RS) & 4096\by4096\by4096 16bit uint & 128 GiB \\ \hline
Forced Isotropic Turbulence\cite{yeung_donzis_sreenivasan_2012} (FI) & 4096\by4096\by4096  16bit uint  & 128 GiB \\ \hline
\end{tabular}
\caption{Datasets used for the measurements.\label{table:datasets}}
\end{table}

\section{Compact VDI Representation}

\begin{figure}
\centering 
\includegraphics[width=\columnwidth]{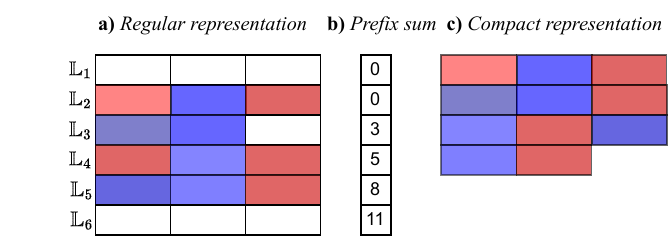}
\caption{Representing the VDI generated in Fig.~\ref{fig:generation} in memory. a) All 18 \superseg{}{} are stored in memory. b) Prefix sum evaluated on (a), which is used to generate the compact representation of the VDI shown in c), storing only the 11 non-empty \superseg{}{}. }
\label{fig:dense}
\end{figure}

VDI generation techniques \cite{Frey} have so far used a regular 3D representation for VDIs. Any lists that pass through empty regions, and therefore do not generate supersegments, or that generate less than \numSupsegs~supersegments, store zeros in the remaining locations as illustrated in Fig.~\ref{fig:dense}a. This regular 3D structure of the VDI improves the performance of raycasting-based rendering as it leads to better memory access patterns. However, for the parallel compositing of VDIs, which requires communication of sub-VDIs between \processings, this generates unnecessary communication overhead.

For the purpose of parallel compositing, we therefore store VDIs in memory using a compact representation. Only those supersegments that are actually generated are stored, as illustrated in Fig.~\ref{fig:dense}c. The difference between the memory required for a compact representation and a regular representation grows when more \processings~are used in a sort-last parallel generation approach. This is because data becoming divided more finely among \processings~leads to sub-VDIs that are increasingly sparse.

The procedure for generating compact VDIs starts similarly to that for regular VDIs. For each ray, a value of \sensitivity~is determined using iterative search (Sec.~\ref{sec:vdi_basics}). No supersegments are actually generated at this stage, but value of \sensitivity~and the number of supersegments that it would generate are both stored in a buffer. Next, a prefix sum is calculated on this buffer, which records for each list \supseglist{i} the total number of supersegments generated by all lists before \supseglist{i} (Fig.~\ref{fig:dense}b). Entry \textit{i} in the prefix buffer, corresponding to list \supseglist{i}, therefore contains the index in the linearized array at which the supersegments of \supseglist{i} start to be stored. All lists can therefore generate their supersegments in parallel, using the value of \sensitivity~determined in the first step. In our implementation, both \sensitivity~search and supersegment generation are parallelized across lists on the GPU, while the prefix sum is calculated on the CPU.

Table~\ref{tab:dense_vs_full} compares the time to generate a compact VDI representation with the time to generate a regular representation, which avoids the prefix-sum computation and the GPU kernel synchronization at the end of the per-ray \sensitivity~search. We find that generating the compact representation is, in fact, faster than generating the regular representation, with the additional time to compute the prefix sum more than amortized by the fact that empty supersegments do not need to be written to memory. 

The compact VDI representation is a form of run-length encoding for the VDI. Though a typical run-length encoding approach would group all neighboring supersegments with the same color and depth values, not only empty ones, it is unlikely in practice for neighboring supersegments to have the same value, unless they are empty. This form of lossless compression is analogous to active-pixel encoding \cite{icet, icetnew} commonly performed in parallel compositing of images, where empty regions in images are compressed using run-lengths.

Although it is possible to render VDIs in their compact representation, this is significantly slower than rendering the regular representation as it offers less optimal memory access patterns (Fig.~\ref{fig:fps_ful_dense}). 
The VDI we finally stream for rendering after parallel compositing is therefore stored in the regular representation, but we use the compact representation during parallel generation and compositing.

\begin{table}
    \centering
\begin{tabular}{c | c | c | c} 
 \hline
 Dataset & Regular & Compact & Memory ratio \\ [0.5ex] 
 \hline\hline
 Kingsnake & 0.42\,s & 0.29\,s & 1:0.12  \\ 
 \hline
 Rayleigh-Taylor & 0.54\,s  & 0.44\,s & 1:0.45 \\
 \hline
 Richtmyer-Meshkov & 0.91\,s & 0.84\,s & 1:0.34 \\
 \hline
\end{tabular}
    \caption{Time in seconds to generate a single \numLists=\fhdres, \numSupsegs=25, VDI stored either using the regular or the compact representation in memory, and the ratio of the memory (regular:compact) required for the representations.}
    \label{tab:dense_vs_full}
\end{table}

\section{Generating a VDI on distributed data}
We propose a method to generate VDIs that represent data distributed across \processings , e.g., compute nodes in a cluster, GPUs within a node, etc. The final VDI represents the entire volume data in the viewport across all \processings~and can be streamed for remote display.

The proposed approach relates to techniques commonly used to generate images from distributed data. We adopt a sort-last parallel rendering strategy \cite{sortlast} in order to achieve scalability with data size and to conform to arbitrary domain decompositions, as, e.g., produced by an \insitu~numerical simulation. Distributed VDI generation therefore begins from a given domain decomposition where each \processing~only stores a part of the overall volume (Fig.~\ref{fig:generation}). Each \processing~then generates a sub-VDI at full viewport resolution for its local data and stores it in the compact representation. The sub-VDIs are composited into a single VDI representing the entire volume data using a direct-send approach~\cite{neumann1933parallel} with compositing load balanced in image space. The compositing stage receives supersegments from all \processings~and combines them to produce a total of up to \numSupsegs~supersegments per list, minimizing loss of detail.

\begin{figure}
\centering 
\includegraphics[width=\columnwidth]{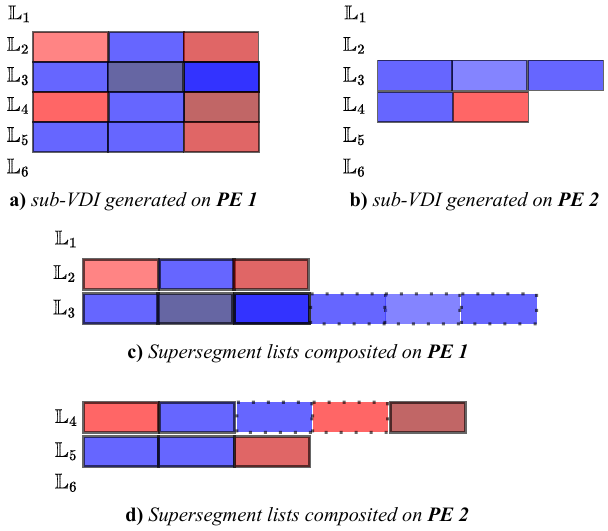}
\caption{The sub-VDIs generated during Phase 1 and the lists composited during Phase 2 for the VDI in Fig.~\ref{fig:generation}.  Panels c) and d) show \subsuperseg{}{}~sorted by their position along the ray. Flat outlines represent \subsuperseg{}{}~received from \processing~1, while dotted outlines represent \subsuperseg{}{} received from \processing~2. These are then composited, producing a maximum of \numSupsegs$=3$ \superseg{}{} per list, as shown in Fig.~\ref{fig:generation}.}
\label{fig:sub_and_comp}
\end{figure}

\subsection{Phase 1: Distributed generation of sub-VDIs} \label{sec:subVDI}
Distributed sub-VDI generation starts from a domain decomposition of the volume data. As is typical for sort-last approaches, no transfer of data between \processings~is required. A VDI corresponding to the full viewport resolution is generated concurrently on each \processing. All \processings~share the camera viewpoint from which rays are cast to generate supersegments. The sub-VDIs are stored in the compact representation. This avoids a linearly growing communication overhead with increasing numbers of \processings, since the sub-VDIs become increasingly sparse for finer data decompositions.

Any given ray in the view frustum can, in general, pass through the domains of multiple \processings. The search for a value of \sensitivity~that would generate a total of up to \numSupsegs~supersegments along the ray would thus require communication between \processings~at each iteration. This would prevent the algorithm from scaling, as \sensitivity~search typically requires several iterations to converge.
We avoid this communication and synchronization by proposing an approximate algorithm that allows each \processing~to act independently during sub-VDI generation. For this, \processings~do not make assumptions about the numbers of supersegments generated on other \processings . Instead, each \processing~independently generates up to \numSupsegs~supersegments (we refer to them as \subsuperseg{}{}) within its own domain. This ensures that the volume rendering integral is never under-resolved for the given budget of \numSupsegs. In Fig.~\ref{fig:generation}, for example, ray 2 passes through the domains of both \processing~1 and \processing~2, but encounters only empty regions on \processing~2. All \numSupsegs$=3$ supersegments are thus generated on \processing~1 (Fig.~\ref{fig:sub_and_comp}). Ray 3, however, generates a full \numSupsegs$=3$ on each \processing, over-resolving the ray with six \subsuperseg{}{}, which are then  reduced during compositing.

The proposed strategy of generating up to \numSupsegs~\subsuperseg{}{}~on each \processing~a ray passes through generates more \subsuperseg{}{} than required. This increases the amount of data to be communicated during compositing. But it eliminates the cost of communicating and synchronizing all \processings~during \sensitivity~search and preserves the best-possible quality for the final VDI. 

\subsection{Phase 2: Parallel compositing of sub-VDIs} \label{sec:compVDI}


At the end of Phase 1, each ray has produced up to \numSupsegs~\subsuperseg{}{} on each \processing~where it intersects the data. These need to be merged to produce a total of up to \numSupsegs~supersegments for each ray. The first step is therefore to bring all \subsuperseg{}{} of a ray from all \processings~onto a single \processing~where they can be composited.

We design an algorithm based on the direct-send approach for compositing sub-images in distributed volume rendering~\cite{neumann1933parallel}. The number of supersegment lists \supseglist{} in the final composited VDI is divided equally among  \processings, with each \processing~responsible for producing composited \supseglist{xy} for the pixels in its part of the image space. In Fig.~\ref{fig:sub_and_comp}, e.g., both \processing~1 and \processing~2 composite three \supseglist{}~each. This balances the load in image space.

Each \processing~sends to all \processings, including itself, the \subsuperseg{}{}~those \processings~are responsible for compositing. Since the amounts of data to be sent to different \processings~differ, we use {\texttt{MPI\_AllToAllv}}, which accounts for variable message lengths. The prefix-sum buffers generated for the sub-VDIs are also transmitted, in chunks corresponding to the image space decomposition, as they are required for reading \subsuperseg{}{}~from the compact representation. Each \processing~then has all the data required for compositing the final supersegments \superseg{}{}~of the lists in its part of the image space.

The first step in merging the \subsuperseg{}{}~is to determine the order in which they lie along the ray. The \subsuperseg{}{}~in any list cannot be assumed to be contiguous. There may be gaps between consecutive \subsuperseg{}{}~when the ray passes through the domain of another \processing . In Fig.~\ref{fig:generation}, e.g., Ray 4 passes through the domain of \processing~2 before returning to \processing~1, generating \subsuperseg{}{} on \processing~2 (see Fig.~\ref{fig:sub_and_comp}) that are to be placed in-between the \subsuperseg{}{} of \processing~1. Within each \processing , however, the \subsuperseg{}{}~are sorted, since they are generated by front-to-back raycasting. Therefore, to iteratively determine the next \subsuperseg{}{}~in a list, the depths of the frontmost \subsuperseg{}{}~from all \processings~are compared, and the \subsuperseg{}{}~with the lowest starting depth is selected as the next along the ray. The merged set of \subsuperseg{}{}~for Ray 4 is produced on \processing~2, which is responsible for compositing \supseglist{4} (Fig.~\ref{fig:sub_and_comp}d).

The process of compositing the merged \subsuperseg{}{}~can be formulated as another supersegment generation task, performed by raycasting through the \subsuperseg{}{}, which are, after all, piecewise constant samples of the original volume. We therefore treat each \subsuperseg{}{}~as a sample along the ray. These samples are raycast through and combined into supersegments \superseg{}{}. Since the \subsuperseg{}{} have different lengths, the process of raycasting through them is analogous to volume raycasting with irregular step size. The opacity obtained from a \subsuperseg{i}{j} is the opacity stored in \subsuperseg{i}{j}, corrected by its length \cite{correction} as:
\begin{equation}
\widetilde{\alpha} = 1 - (1-\alpha)^l
\end{equation}
where $\widetilde{\alpha}$ is the adjusted opacity, $\alpha$ is the opacity stored in \subsuperseg{i}{j}, and $l$ is the length of \subsuperseg{i}{j}.
Empty spaces between \subsuperseg{}{} are treated as transparent samples. At each sample, the \subsuperseg{}{} can either be merged with the previous supersegment, or it can begin a new one. This is again determined using the criterion \criteria~(Eq.~\ref{eq:termination}) and requires finding another \sensitivity~that leads to the generation of \numSupsegs~supersegments in total. This is again done using per-ray iterative \sensitivity~search (Sec.~\ref{sec:vdi_basics}), which requires multiple passes through the \subsuperseg{}{}. The \subsuperseg{}{}~in \supseglist{4} in Fig.~\ref{fig:sub_and_comp}d are, e.g., combined to produce the \superseg{}{} depicted in Fig.~\ref{fig:generation}. Detailed pseudocode is included in the Supplement.

\begin{figure}
\centering 
\includegraphics[width=0.9\columnwidth]{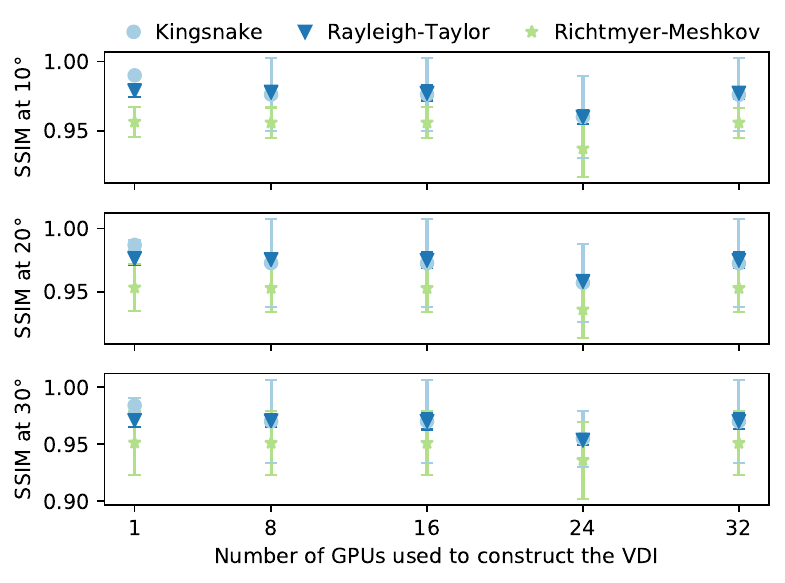}
\caption{SSIM (mean $\pm$ min-max range across four viewpoints of VDI generation) with respect to ground truth DVR for the rendering of VDIs generated using varying numbers of Nvidia A100 GPUs. VDIs generated on 8, 16, 24 and, 32 GPUs are composited using the presented compositing algorithm (Sec.~\ref{sec:compVDI}), while VDIs generated on 1 GPU do not undergo compositing. All VDIs are of  resolution \numLists=1920\by1080 with \numSupsegs=25 for three datasets (symbols, top legend) and three different \viewnew~(panels).}
\label{fig:compositing_q}
\end{figure}

\begin{figure*}
\centering
\begin{subfigure}{.32\textwidth}
  \centering
  \includegraphics[width=\linewidth]{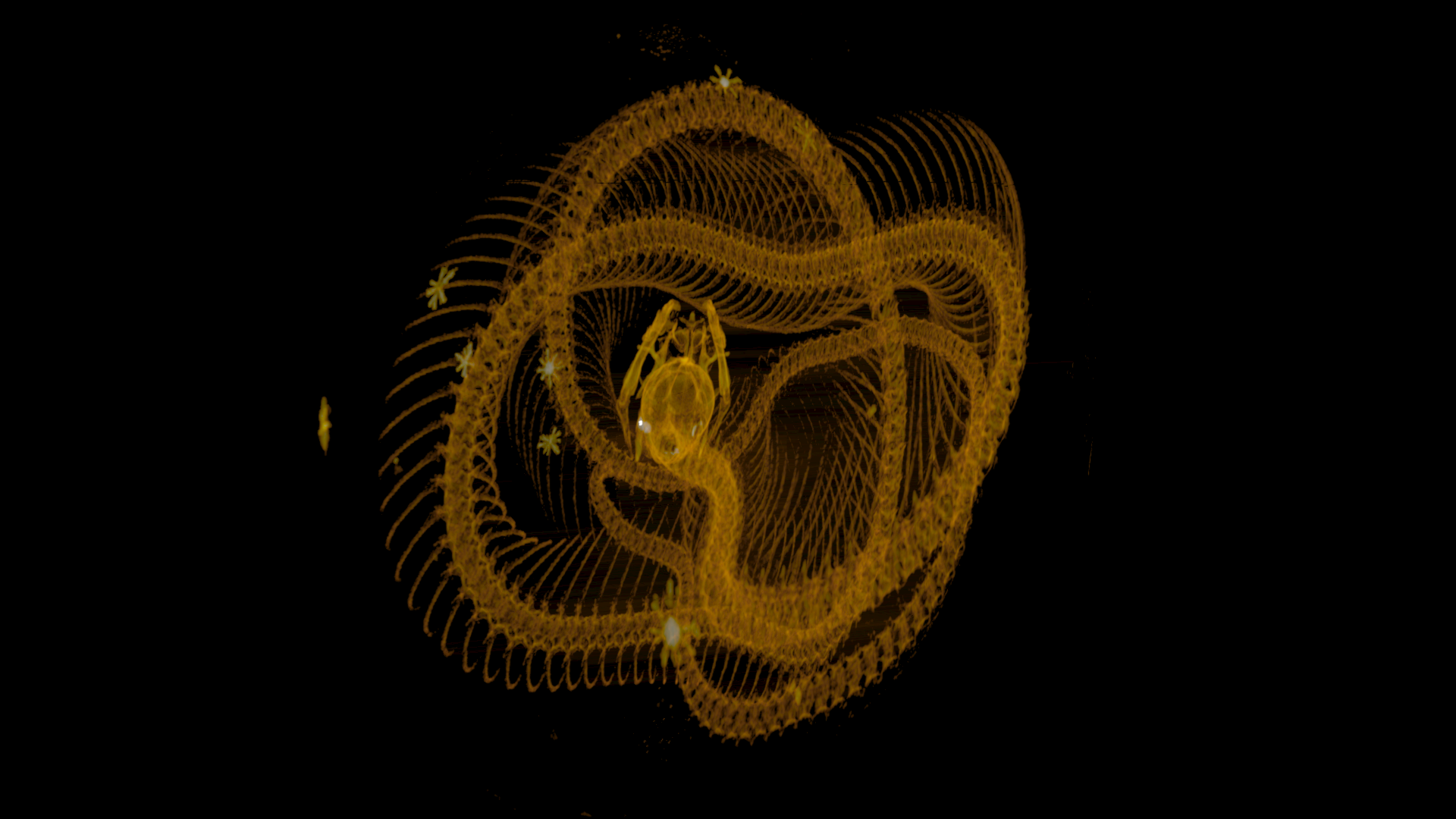}
  \caption{Kingsnake, SSIM: 0.970.}
\end{subfigure}%
~
\begin{subfigure}{.32\textwidth}
  \centering
  \includegraphics[width=\linewidth]{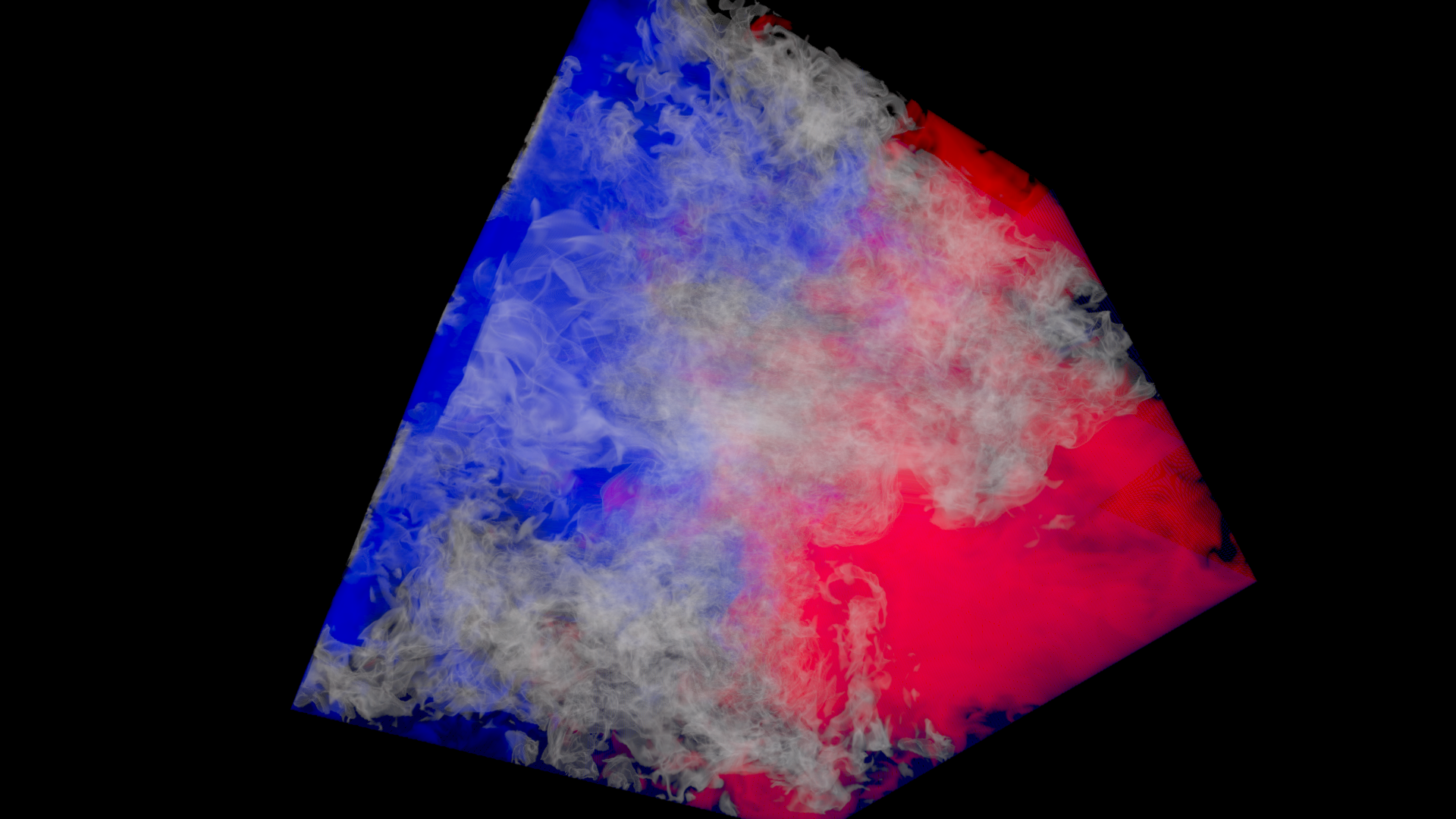}
  \caption{Rayleigh-Taylor, SSIM: 0.974.}  
\end{subfigure}
~
\begin{subfigure}{.32\textwidth}
  \centering
  \includegraphics[width=\linewidth]{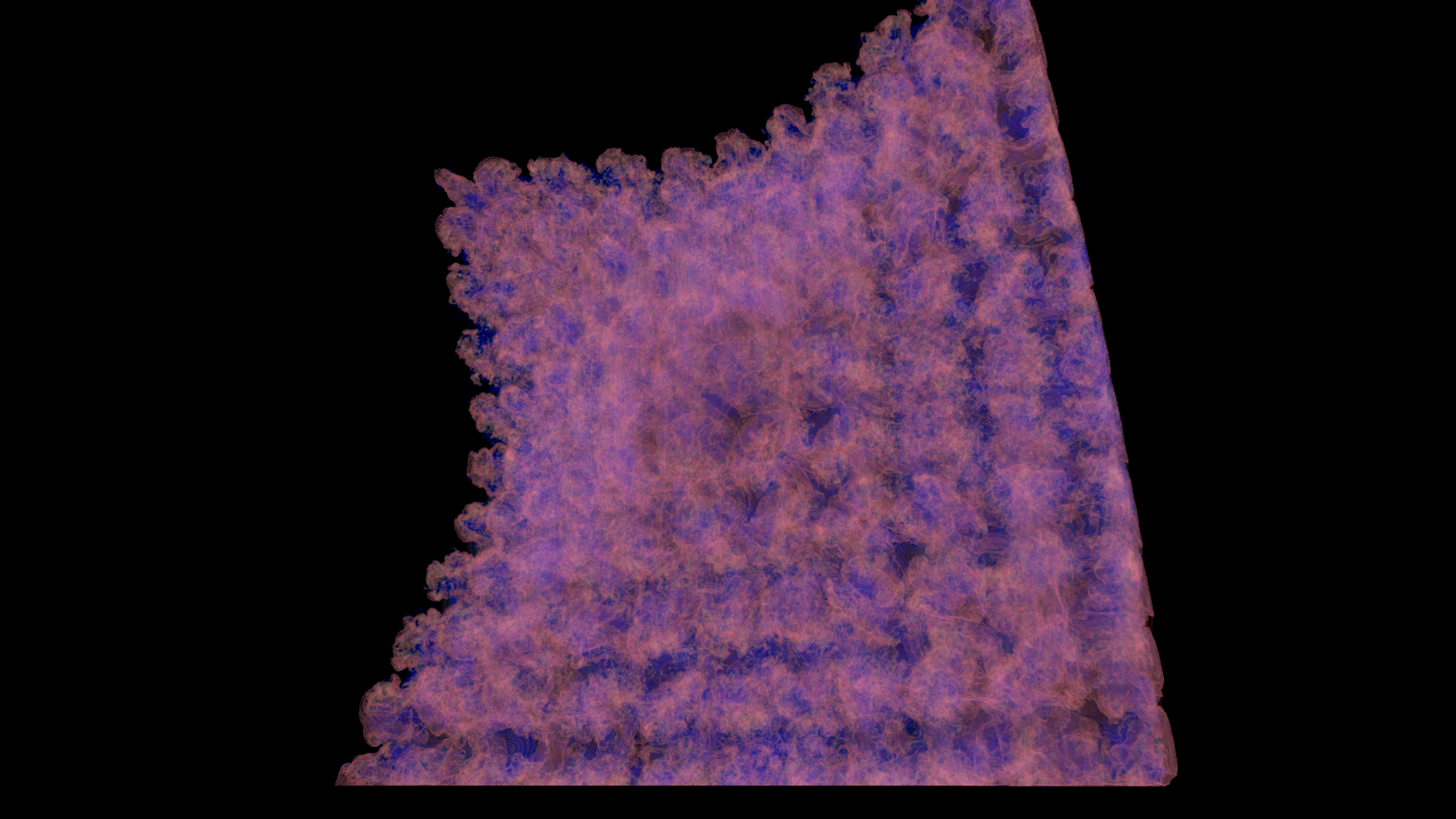}
  \caption{Richtmyer-Meshkov, SSIM: 0.949.}
\end{subfigure}

\caption{Visual illustration of VDI rendering quality. VDIs generated on 32 GPUs are rendered at 20\textdegree~from the viewpoint of generation. SSIM values computed w.r.t.~ground-truth direct volume rendering (DVR) at the same viewpoint (see Supplement for images).}
\label{fig:single_image_visual}
\end{figure*}

While parallel compositing approaches like binary-swap \cite{ma1993data} and radix-k \cite{peterka2009configurable} typically outperform the direct-send for image compositing, we choose the latter here as it requires only a single stage of compositing \subsuperseg{}{}~into \superseg{}{}. The proposed sort-last approach of compositing \subsuperseg{}{}~produces different \superseg{}{}~depending on the number of \processings~the data is divided over. To evaluate the accuracy, we generate VDIs on multiple GPUs~and compare the quality of the rendering with a VDI generated on a single GPU, where compositing is not required. VDIs are generated from four viewpoints (\vieworig) representing 90\textdegree~rotations of the dataset in camera space. They are then rendered at different viewpoint deviations (\viewnew) about the viewpoint of generation, and quality is compared to ground truth direct volume rendering (DVR). Figure~\ref{fig:compositing_q} presents the results with image similarity measured using the SSIM \cite{ssim} metric, depicting the mean and the range (max-min) across the four \vieworig. While there is a marginal decline in SSIM values, we observe that VDI rendering quality remains similarly high for VDIs generated on multiple GPUs as for a VDI generated on a single GPU. Transfer functions are depicted in Fig.~\ref{fig:single_image_visual} and images are included in the Supplement for visual comparison.

\begin{figure}
\centering 
\includegraphics[width=\columnwidth]{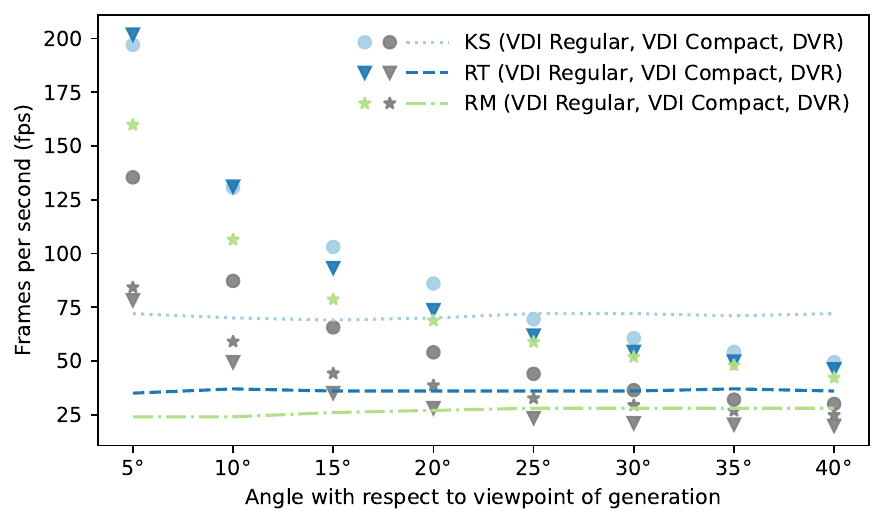}
\caption{Performance comparison (in fps) between DVR (lines) and rendering \numLists=1920\by1080, \numSupsegs=25, VDIs stored in either regular (colors) or compact (grayscales) representation, at various angles about the viewpoint of generation. Datasets (Kingsnake (KS), Rayleigh-Taylor (RT), Richtmyer-Meshkov (RM)) are represented by symbols/line styles, see inset legend.}
\label{fig:fps_ful_dense}
\end{figure}

Figure~\ref{fig:fps_ful_dense} compares the rendering frame rates of VDIs stored in regular and compact representation with the DVR baseline at different \viewnew~about \vieworig. The regular representation renders faster due to its simpler memory-access patterns. The task of converting from compact to regular representation cannot be handled by the display client as it needs to remain responsive for user interaction. Therefore, while the sub-VDIs communicated for compositing used the compact representation to reduce communication overhead, the supersegment lists produced by compositing use the regular representation. The composited lists from all \processings~are combined onto the root \processing~by {\texttt{MPI\_Gather}}, from where they are be streamed for (remote) display, potentially after lossless compression.

\subsection{Handling Non-Convex Data Decompositions}\label{sec:nonconvex}
A key feature of the above compositing method is that it can handle non-convex domain decompositions and therefore work with any application-given data distribution.

A non-convex domain decomposition is one where a ray can intersect the boundary of the domain of a \processing~in more than two points. Such decompositions occur, e.g., in numerical simulations in complex-shaped simulation domains, where the domain decomposition balances the computations in each sub-domain and the communication volume between \processings~\cite{incardona2019openfpm}, not necessarily producing an equal division of data among \processings. Such situations are challenging for distributed volume rendering, due to the non-commutativity of the {\texttt{over}} operator \cite{over}:
\begin{equation}
    a \texttt{ over } b \,\,\, \neq \,\,\,  b \texttt{ over } a .
\end{equation}
This implies that in non-convex domain decompositions, volume rendering cannot composite color across disjoint segments of a ray without requiring communication or synchronization between the \processings, or redistribution of the volume data.

Our method avoids this problem by generating \subsuperseg{}{} that store world-space front and back depths along the ray. A \subsuperseg{}{} necessarily terminates when the ray leaves the domain of a \processing. Since \subsuperseg{}{} are ordered by their depth during compositing, {\texttt{over}} operations are done in the correct order. The \subsuperseg{}{} along a ray can therefore be generated in parallel without synchronization or communication between the \processings.

The present approach includes non-convex distributed volume rendering as a limit case: when generating only a single \subsuperseg{}{} per sub-domain intersection, the compositing algorithm can, in addition to placing the supersegments in correct order, also perform {\texttt{over}}-operator compositing along the supersegment lists. This effectively performs volume rendering, creating a flat image on a non-convex domain decomposition without requiring synchronization or communication between \processings.

\section{Implementation}

We implemented the algorithms described in the previous sections in the open-source rendering framework \scenery. Both sub-VDI generation and VDI compositing are implemented as compute shaders in the Vulkan API. For work distribution, a local work-group size of 16\by16 is used, i.e., the screen space is divided into 2D blocks of that size. During raycasting, each ray within a block corresponds to a thread on the GPU, and to a single pixel on screen. 

The final VDI generated is compressed using LZ4, which we found to provide faster and better compression than Snappy and ZSTD, before streaming. For a \numLists=1920\by1080, \numSupsegs=25, VDI this produces $\approx$325\,MiB of data, while the corresponding uncompressed VDI in regular 3D representation would be $\approx$1.2 GiB.

The source code is available under the open-source BSD license at: \githuburl.

\section{Benchmarks and Evaluation}

We evaluate parallel VDI generation on the real-world datasets from Table~\ref{table:datasets}, measuring rendering performance and quality.

\begin{figure}
\centering 
\includegraphics[width=\columnwidth]{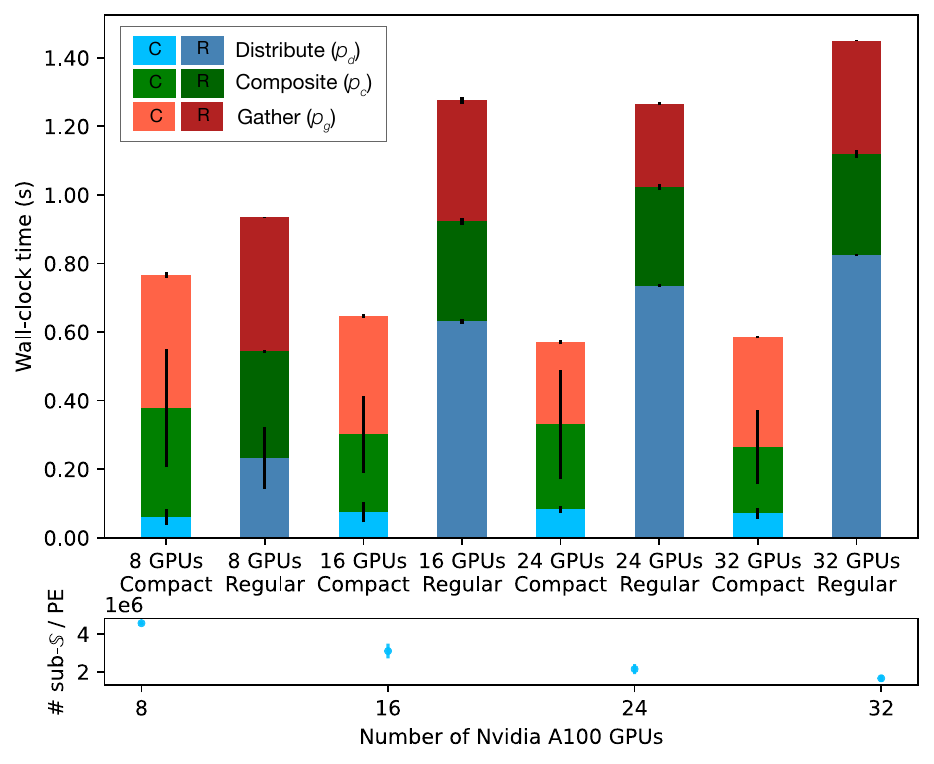}
\caption{Runtime performance of the three stages of parallel compositing for VDIs in both compact (C, light colors) and regular (R, dark colors) representation. Mean $\pm$ standard deviation (error bars) are reported over a 360\textdegree~camera orbit (top panel). The individual parts are: distribution of \subsuperseg{}{}~($p_d$, blue), compositing into \superseg{}{}~($p_c$, green), final gathering of \superseg{}{}~($p_g$, red). The bottom panel shows the average number of \subsuperseg{}{}~generated by a \processing.}
\label{fig:compositing_prof}
\end{figure}

We profile the performance of the individual steps of parallel compositing (Sec.~\ref{sec:compVDI}) in order to quantify the benefits of using the proposed compact VDI representation. Figure~\ref{fig:compositing_prof} plots, for both compact and the regular representation, the timings of the three stages of parallel compositing: the distribution of \subsuperseg{}{}~among \processings~($p_d$), the actual compositing into \superseg{}{} ($p_c$), and the final gather at the root \processing~($p_g$). An \texttt{MPI\_Barrier} is placed before all MPI calls for profiling. Measurements are reported as mean $\pm$ standard-deviation over 144 iterations with the camera rotating 5\textdegree~about the dataset every second iteration for a full 360\textdegree~orbit. For $p_d$ and $p_c$, timings are averaged across \processings~at every iteration, since the \processings~can independently proceed with the next step; $p_g$ is recorded at the root \processing. VDIs of the FI dataset (Table~\ref{table:datasets}) are used in full HD (\fhdres) viewport resolution with \numSupsegs=25.

\begin{figure}
\centering 
\includegraphics[width=\columnwidth]{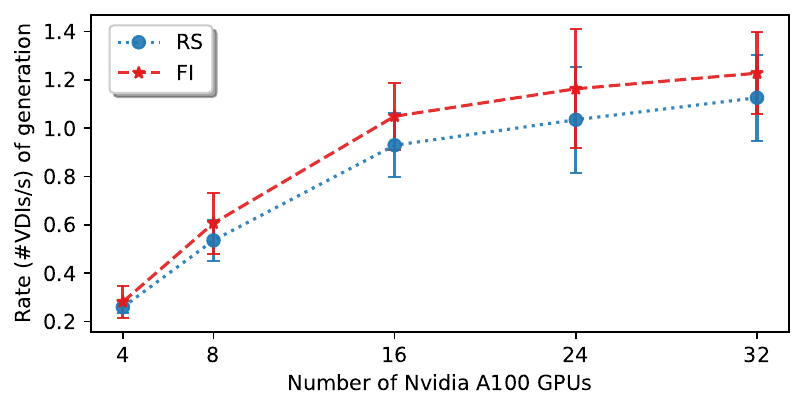}
\caption{Number of VDIs generated per second (mean $\pm$ standard deviation over a 360\textdegree~camera orbit) for different numbers of Nvidia A100 GPUs, \numLists=\fhdres, \numSupsegs=25, for the RS and FI datasets of 128\,GiB each (see Table \ref{table:datasets}).}
\label{fig:overall_vdi}
\end{figure}

\begin{figure*}
    \centering
    \begin{subfigure}{0.48\textwidth}
        \centering
        \includegraphics[width=\linewidth]{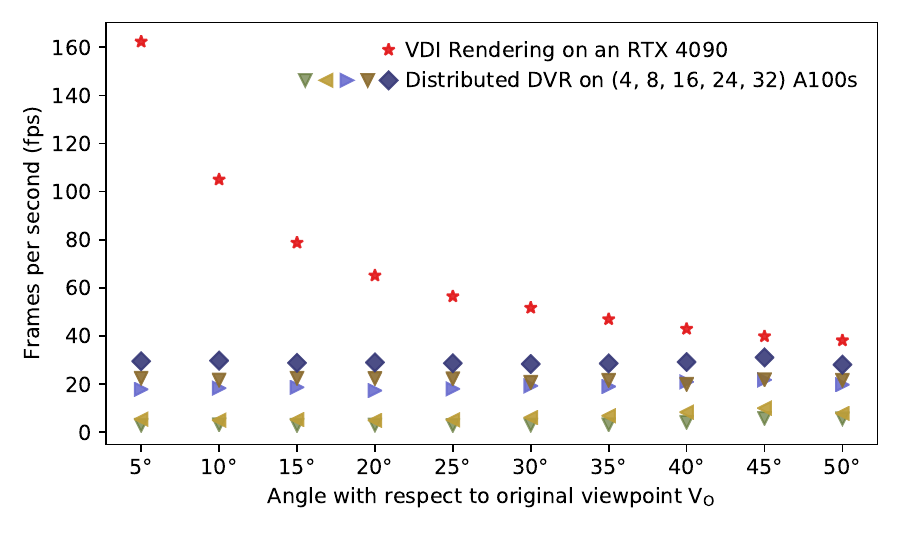}
        \caption{Forced Isotropic Turbulence (FI) dataset, 128\,GiB.}
        \label{fig:benchmark_isotropic}
    \end{subfigure}
    \begin{subfigure}{0.48\textwidth}
        \centering
        \includegraphics[width=\linewidth]{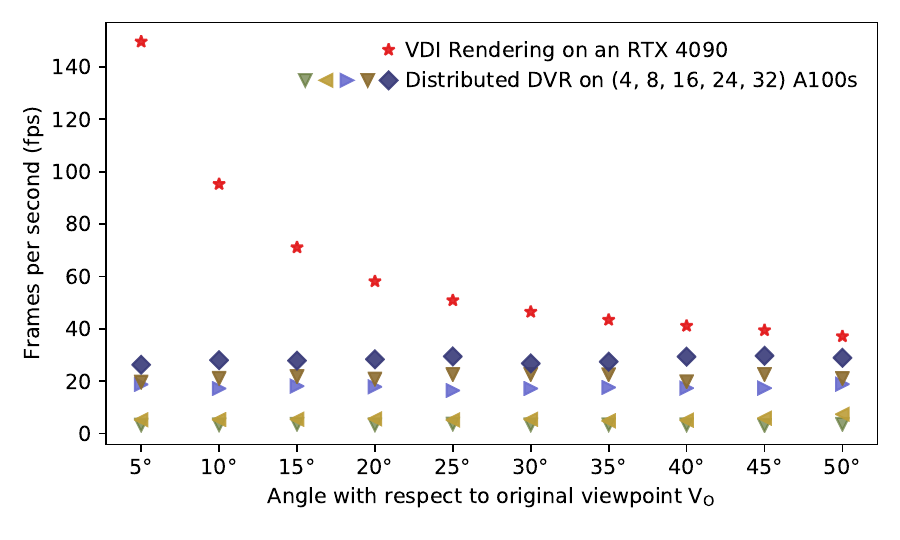}
        \caption{Rotating Stratified Turbulence (RS) dataset, 128\,GiB.}
        \label{fig:benchmark_rotstrat}
    \end{subfigure}
    \caption{VDI rendering frame rates on a display client with one Nvidia RTX 4090 (stars),
    averaged over four viewpoints at 90\textdegree~from each other, compared with distributed DVR on the cluster using IceT for sort-last parallel compositing on varying numbers of Nvidia A100s (other symbols, see inset legend); resolution \fhdres~with \numSupsegs=25.}
    \label{fig:vdi_vs_distr_volume}
\end{figure*}

Using the compact VDI representation during compositing is faster in all tested cases and provides better scalability with increasing numbers of GPUs. On 32 GPUs, the entire parallel compositing process, i.e.~$(p_d + p_c + p_g)$, for the compact representation requires 40\% of the time of the regular representation. As expected, the average number of \subsuperseg{}{}~generated per \processing~decreases when using more \processings, as the sub-VDIs become increasingly sparse. This amplifies the advantage of the compact representation, reducing $p_d$ by about 90\% on 32 GPUs.

\begin{figure} 
        \centering
        \includegraphics[width=\columnwidth]{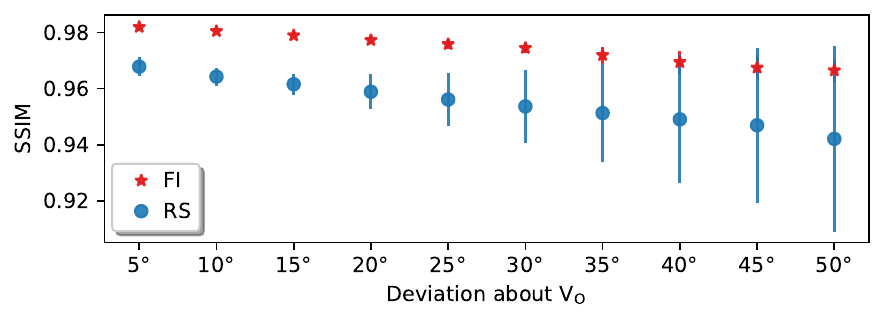}
        \caption{SSIM image similarity between VDI rendering and DVR for different \viewnew. Mean and min-max range (error bars) are reported over four different \vieworig~for two datasets (RS: circles, FI: stars).}
    \label{fig:ssim_vdi}
\end{figure}

\begin{figure*}
    \centering
    \includegraphics[width=\textwidth]{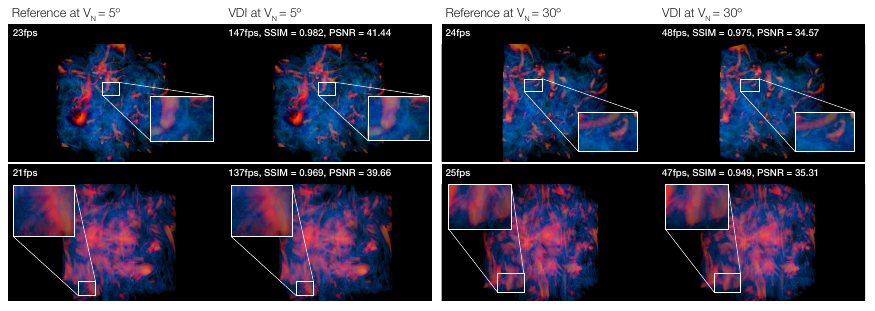}
    \caption{Visual comparison of VDI rendering quality with distributed DVR (``Reference'') for the Forced Isotropic Turbulence (top) and Rotating Stratified Turbulence (bottom) datasets with a multi-hue transfer function. Image quality metrics are computed w.r.t.~the DVR image.}
    \label{fig:visual_comp}
\end{figure*}

The reduction in the number of \subsuperseg{}{}~per \processing~also reduces the compositing time ($p_c$) to $0.6\times$ on 32 GPUs, relative to 8 GPUs. VDI sparsity, however, depends on \vieworig, resulting in the higher standard-deviation for the compact representation. For the regular representation, the number of \subsuperseg{}{}~per \processing~remains constant, leading to constant compositing times. The communication volume for the final gather ($p_g$) is independent of the number of \processings~and equal to the size of the final VDI. The observed variability is thus attributed to background load fluctuations on the benchmark machine. Eventually, $p_d$ limits scalability, as it increases ($1.4\times$ on 32 GPUs relative to 8 GPUs) even for the compact representation. 

Figure~\ref{fig:overall_vdi} shows the overall rate of VDI generation for the 128\,GiB RS \cite{rotstrat} and FI \cite{yeung_donzis_sreenivasan_2012} datasets, starting at 4 GPUs which is the minimum required for this data size (Table~\ref{table:datasets}). Again, the camera performs a 360\textdegree~orbit of the data at 5\textdegree~increments with two iterations at each viewpoint.

At small numbers of GPUs, sub-VDI generation dominates the overall VDI generation time. As the number of GPUs increases, sub-VDI times decrease (strong scaling). From 4 to 8 GPUs, VDI generation rates therefore increase by $2.2\times$ and $2.1\times$ for the FI and RS datasets, respectively. As sub-VDI times decrease further, the communication overhead of parallel compositing begins to dominate, and speed-ups reduce to $1.75\times$ from 8 to 16 GPUs and $1.2\times$ from 16 to 32 GPUs. This is typical for strong scaling and is also observed in distributed volume rendering \cite{icetnew}. Our implementation overlaps sub-VDI generation with the final gather  of the previous VDI, effectively hiding some of the sub-VDI generation time. This increases VDI generation rates, but decreases the measured parallel efficiency. Overall, about one VDI is generated per second on 32 GPUs.
The results are similar for other values of \numSupsegs. VDI generation rates increase by $\approx 15\%$ for \numSupsegs=20 and decrease by $\approx 9\%$ for \numSupsegs=30, relative to the \numSupsegs=25 plotted here (see Supplement).

Once the final VDI is generated, it can be streamed for interactive remote rendering. We therefore compare VDI rendering on the display client with distributed DVR on the cluster. We implement distributed DVR using IceT \cite{icet}, a commonly used library for sort-last parallel compositing, as also used by ParaView \cite{paraview} and VisIt \cite{VisIt}. To ensure optimal compositing performance, we recompile IceT with support for the recent CUDA image compression extension \cite{icetnew}. The sort-last parallel compositing strategy is chosen using the \texttt{ICET\_SINGLE\_IMAGE\_STRATEGY\_AUTOMATIC} option, which automatically selects between the radix-k \cite{peterka2009configurable}, binary swap \cite{ma1993data}, and binary tree \cite{icet} methods, based on the number of processes. This is the default setting in IceT.

VDIs are rendered at different viewpoint deviations (\viewnew) about the viewpoint of generation (\vieworig) and compared to distributed DVR from the same viewpoint. Results are reported at four different \vieworig~at 90\textdegree~rotation increments around the data. Transfer functions with multi-hue colormaps are used to present a challenging test case for the VDI. Volume raycasting, both for DVR and VDI generation, used an emission-absorption illumination model.

Figure~\ref{fig:vdi_vs_distr_volume} reports rendering frame rates averaged over the four \vieworig. For distributed DVR, frame rates are  limited by volume raycasting with parallel compositing adding an overhead of $\approx 12$\,ms per frame on 32 GPUs. Close to \vieworig, VDI frame rates are significantly higher than distributed DVR, achieving a speed-up of $5.5\times$ at \viewnew=5\textdegree~for both datasets. For larger deviations, VDI frame rates reduce due to the anisotropic view-dependent shape of the VDI. But they remain at least $1.3\times$ better than for distributed DVR even at \viewnew=50\textdegree.

Finally, we compare the quality of the images generated by VDI rendering with those from DVR. VDIs generated on 32 GPUs at the same four \vieworig~are rendered on a remote workstation and compared with distributed DVR on 32 GPUs. Figure~\ref{fig:ssim_vdi} reports the  SSIM (Structural Similarity Index Measure) \cite{ssim} between the VDI rendering result and DVR, with 1.0 indicating identical images. VDI rendering provides high-quality approximations to DVR for both the RS and FI datasets. As expected, SSIM reduces with increasing viewpoint deviation, but overall remains high even at 50\textdegree. The results are similar when measuring rendering quality in terms of PSNR (Peak Signal-to-Noise Ratio, see Supplement). Visual comparisons are given in Fig.~\ref{fig:visual_comp} with full-resolution images available in the Supplement, along with screencast videos of interactive rendering on the FI and RS datasets.

\section{Conclusions}
We have presented a distributed generation and parallel compositing approach for Volumetric Depth Images (VDIs), enabling responsive interactive visualization of large distributed volumes at consistently high frame rates. We improved the scalability of parallel compositing using a compact memory layout for VDIs that stores only non-empty supersegments, analogous to active-pixel encoding in parallel image compositing \cite{icet, icetnew}. We found that this significantly reduced communication overhead during compositing (Fig.~\ref{fig:compositing_prof}) and that VDIs in the compact representation were slightly faster to generate than those represented at regular 3D resolution (Table~\ref{tab:dense_vs_full}). After compositing, however, the VDIs were represented at regular 3D resolution in memory, as we found this to allow for significantly faster rendering (Fig.~\ref{fig:fps_ful_dense}). 

We followed a sort-last approach where sub-supersegments received for compositing were treated as samples of varying length and $\alpha$-composited along the ray to form the final supersegments. This produced accurate VDI approximations (Fig.~\ref{fig:compositing_q}) even on large datasets (Figs.~\ref{fig:ssim_vdi}, \ref{fig:visual_comp}) and allowed us to generate sub-VDIs in parallel without any communication between processing elements. This is particularly beneficial in combination with the automated \sensitivity~parameter optimization, where synchronization would otherwise occur at each iteration and for each ray. Overall speed-ups for VDI generation (Fig.~\ref{fig:overall_vdi}) reduced from linear to sub-linear with increasing GPU counts as soon as compositing times became dominant, as expected for strong scaling.

The proposed distributed generation approach enabled us to test the VDI on larger data than what was possible in previous works. We found that VDI rendering on a display client was about $5.5\times$ faster than distributed DVR near the viewpoint of generation and remained faster also at large viewpoint deviations (Fig.~\ref{fig:vdi_vs_distr_volume}). The rendered images were almost identical, particularly close to the viewpoint of generation (Figs.~\ref{fig:ssim_vdi}, \ref{fig:visual_comp}).

Overall, VDIs were generated in less than one second when using 32 GPUs for a 128\,GiB datasets (Fig.~\ref{fig:overall_vdi}). These generation times are comparable to or lower than the iteration times of typical distributed numerical simulations. We therefore envision distributed VDI generation finding applications in interactive \insitu~visualization of numerical simulations. This also includes simulations on unstructured grids, as VDIs can be generated for any volume discretization that can be raycast. While this will influence sub-VDI generation times, as it would for any rendering, it does not change the downstream parallel compositing presented here. We further see potential applications with the Cinema database \cite{cinema} for exploratory post-hoc visualization, where distributed VDIs could be used to reduce the size of the database by reducing the need for images generated from different viewpoints. 

The present parallel approach for compositing \subsuperseg{}{}~into \superseg{}{}~maintains scalability with increasing \processings, reducing compositing times (Fig.~\ref{fig:compositing_prof}) despite an increase in the overall number of \subsuperseg{}{}~due to our strategy of generating up to \numSupsegs~\subsuperseg{}{}~per list on each \processing. On the other hand, a limitation is that it is susceptible to load imbalance due to variation in the number of \subsuperseg{}{}~across lists. 
This is because we equi-distribute lists among \processings. Future optimizations could explore alternate strategies for distributing lists that balance the \subsuperseg{}{}~distribution among \processings.
Whether the gain in load balance amortizes the additional global communication required to do so, however, remains to be seen. In addition, the proposed approach inherits limitations of the VDI representation. Since VDIs store transfer-function classified color and opacity, they do not support interactive transfer-function modification. Also, gradients of the volume data cannot be calculated from the VDI, precluding the use of directional illumination effects, such as specular lighting.

Notwithstanding these limitations, we believe that the methods and algorithms presented here are key in enabling the use of view-dependent volume representations, such as the VDI, for interactive visualization of large distributed volume data.

\section*{Acknowledgments}{
This work was supported by the Center for Scalable Data Analytics and Artificial Intelligence (ScaDS.AI) Dresden/Leipzig, by the German Federal Ministry of Education and Research (BMBF) in the joint project ``6G-life'' (ID 16KISK001K), and by the Center for Advanced Systems Understanding (CASUS) financed by the BMBF and the Saxon Ministry for Science, Culture and Tourism (SMWK) with tax funds on the basis of the budget approved by the Saxon State Parliament. We thank the Center for Information Services and High Performance Computing (ZIH) of TU Dresden for providing their facilities for the benchmarks. We thank the  University of Texas High-Resolution X-ray CT Facility (UTCT) for the Kingsnake dataset.
}

\bibliographystyle{eg-alpha-doi} 
\bibliography{egbibsample}   

\pagebreak

\include{supplementary}

\end{document}

%% file: supplementary.tex
\renewcommand{\theequation}{S\arabic{equation}}
\renewcommand{\thetable}{S\arabic{table}}
\renewcommand{\thefigure}{S\arabic{figure}}
\renewcommand{\thealgocf}{S\arabic{algocf}}
\renewcommand{\thesection}{S\arabic{section}}
\setcounter{equation}{0}
\setcounter{table}{0}
\setcounter{figure}{0}
\setcounter{algocf}{0}
\setcounter{section}{0}

{\LARGE Supplementary Material}
\section{Compositing sub-supersegments \subsuperseg{}{} to form supersegments \superseg{}{}}

As mentioned in the main text (Sec. 5.2), the process for compositing \subsuperseg{}{}~received from multiple \processings~to generate a list of supersegments \superseg{}{}~models the \subsuperseg{}{} as discrete samples to be raycast over, combining them with $\alpha$-compositing, and running an iterative \sensitivity~search until up to \numSupsegs~are generated. Algorithm~\ref{alg:combining} shows the corresponding pseudocode for combining \subsuperseg{}{} into \superseg{}{} for a given value of \sensitivity, i.e., this algorithm is iterated over during the \sensitivity~search, and then finally used to generate the supersegments. Notations are consistent with those introduced in the main text.

\section{Impact of \numSupsegs~on VDI Generation Rates}

\begin{figure}
\centering
\begin{subfigure}{.47\textwidth}
  \centering
  \includegraphics[width=\linewidth]{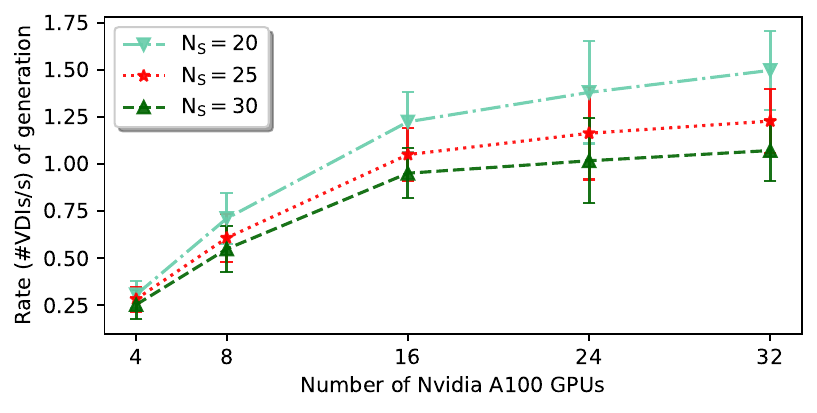}
  \caption{Forced Isotropic Turbulence (FI).}
\end{subfigure}

\begin{subfigure}{.47\textwidth}
  \centering
  \includegraphics[width=\linewidth]{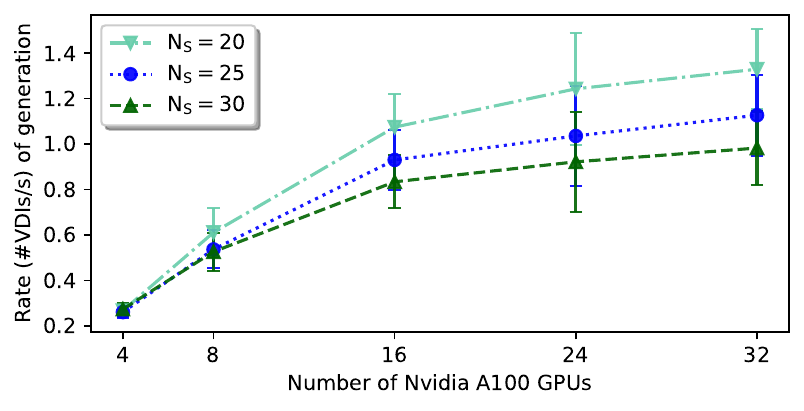}
  \caption{Rotating Stratified Turbulence (RS).}  
\end{subfigure}
\caption{Number of VDIs generated per second (mean $\pm$ standard deviation over a 360\textdegree~camera orbit) for different numbers of Nvidia A100 GPUs and different values of \numSupsegs. \numLists=\fhdres~VDIs are generated for the Forced Isotropic Turbulence (FI) (a) and Rotating Stratified Turbulence (RS) (b)  datasets of 128\,GiB each.}
\label{fig:different_ns}
\end{figure}

To supplement the VDI generation rates for different numbers of GPUs reported in the main text for \numSupsegs=25, Figure~\ref{fig:different_ns} additionally plots generation rates for \numSupsegs=20 and \numSupsegs=30 for both the Forced Isotropic Turbulence (FI) and Rotating Stratified Turbulence (RS) datasets. As in the main text, means and standard deviations are plotted over a 360\textdegree~camera orbit with the camera rotating in 5\textdegree~increments and two iterations at each viewpoint. Trends are similar at all values of \numSupsegs, with linear speed-ups for lower GPU counts reducing to sub-linear at higher counts. As expected, VDIs with smaller \numSupsegs~are faster to generate.

\section{Quality of VDI Rendering in PSNR}
\begin{figure}[!h]
        \centering
        \includegraphics[width=\columnwidth]{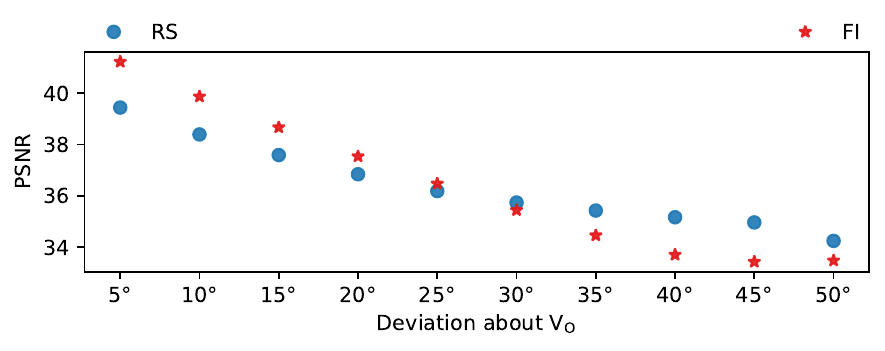}
        \caption{PSNR image similarity between VDI rendering and DVR for different values of \viewnew. Mean values are reported over four different \vieworig~for two datasets (RS: circles, FI: stars).}
    \label{fig:psnr_vdi}
\end{figure}

Figure 10 in the main text compares the quality of VDI rendering with distributed direct volume rendering (DVR) using the SSIM metric. Figure~\ref{fig:psnr_vdi} reports the same quality comparison using the Peak Signal-to-Noise Ratio (PSNR) as a metric. As in the main text, a VDI generated on 32 GPUs is rendered and compared with DVR on 32 GPUs. Mean values are reported over four viewpoints of VDI generation (\vieworig) at 90\textdegree~rotations about the dataset.

\begin{algorithm}  
\DontPrintSemicolon
\SetAlgoLined
\caption{Combining sub-supersegments into supersegments}          
\label{alg:combining}                          
\begin{algorithmic} [1]                   
\STATE \tcc{Combining sub-supersegments along a ray into supersegments for a given \sensitivity}
\STATE supersegmentIsOpen $\gets$ False\; 
\STATE \front{}{}, \back{}{} $ \gets 0 $\;
\STATE \back{}{}$_t$ $\gets 0$ \;
\STATE C(\superseg{}{}), $\alpha$(\superseg{}{}) $\gets$ 0\;
\STATE transparent $\gets$ False\;
\;
\tcc{The front element of all lists is initialized to the index of the first element}
\STATE frontIndex$[0 \ldots n]$ $\gets 1$\; 

\WHILE{!samplesComplete}
\STATE C(\subsuperseg{}{}), $\alpha$(\subsuperseg{}{}), f(\subsuperseg{}{}), b(\subsuperseg{}{}), p $\gets$ findNextSub(frontIndex[]) \;
\IF{p = -1}
\STATE samplesComplete $\gets$ True \;
\ENDIF
\STATE l $\gets$ distance(f(\subsuperseg{}{}), b(\subsuperseg{}{})) \;
\STATE $\widetilde{\alpha}$(\subsuperseg{}{}) = 1 - (1-$\alpha$(\subsuperseg{}{}))$^l$ \;

\IF{supersegmentIsOpen}
\IF{f(\subsuperseg{}{}) $>$ \back{}{}}
\STATE transparent $\gets$ True \;
\STATE C(\subsuperseg{}{}), $\alpha$(\subsuperseg{}{}) $\gets 0$ \; 
\STATE b(\subsuperseg{}{}) $\gets$ f(\subsuperseg{}{}) \;
\STATE f(\subsuperseg{}{}) $\gets$ \back{}{} \;
\ENDIF
\STATE $\widetilde{C(\superseg{}{})} \gets$ $\frac{C(\superseg{}{})}{\alpha(\superseg{}{})}$) \;
\STATE l$_{s} \gets$ distance(\front{}{}, \back{}{}) \;
\STATE $\widetilde{\alpha(\superseg{}{})} = 1 - (1-\alpha(\superseg{}{}))^{l_{s}}$

\IF{\sensitivity $< || \widetilde{C(\mathbb{S})}\widetilde{\alpha(\mathbb{S})} - C(\subsuperseg{}{})\alpha(\subsuperseg{}{}) ||_2$}
\STATE newSupersegment $\gets$ True \;
\ENDIF

\IF{newSupersegment $||$ samplesComplete}
\tcc{Closing \superseg{}{}. If \sensitivity~was the final value determined by iterative search, store \subsuperseg{}{}}
\STATE numTerminations $\gets$ numTerminations + 1 \;

\ELSE
\STATE C(\superseg{}{}) $\gets$ C(\superseg{}{}) + (1 - $\alpha$(\superseg{}{})) * C(\subsuperseg{}{}) * $\alpha$(\subsuperseg{}{}) \;
\STATE $\alpha$(\superseg{}{}) $\gets$ $\alpha$(\superseg{}{}) + (1 - $\alpha$(\superseg{}{})) * $\alpha$(\subsuperseg{}{}) \;
\STATE \back{}{} $ \gets$ b(\subsuperseg{}{})\;
\IF{!transparentSample}
\STATE \back{}{}$_t$ $\gets$ b(\subsuperseg{}{})\;
\ENDIF
\IF{!supersegmentIsOpen \& !transparent}
\STATE supersegmentIsOpen $\gets$ True \;
\STATE \front{}{} $\gets$  f(\subsuperseg{}{}) \;
\STATE \back{}{}, \back{}{}$_t$ $ \gets$ b(\subsuperseg{}{})\;
\STATE C(\superseg{}{}) $\gets$ C(\subsuperseg{}{}) * $\alpha$(\subsuperseg{}{}) \;
\STATE $\alpha$(\superseg{}{}) $\gets$ $\alpha$(\subsuperseg{}{}) \;
\ENDIF
\ENDIF
\ENDIF
\algstore{combining}
\end{algorithmic}
\end{algorithm}

\begin{algorithm}                     
\begin{algorithmic} [1]                   
\algrestore{combining}
\IF{p $\neq$ -1 \& !transparent}
\tcc{Increment the front index of the process whose supersegment was selected}
\STATE frontIndex$[p]\gets$frontIndex$[p]+1$\;
\ENDIF
\ENDWHILE
\end{algorithmic}
\end{algorithm}

